\documentclass[prc,aps,nofootinbib,twocolumn,superscriptaddress]{revtex4-1}   
\usepackage{epsfig}  
\usepackage{graphicx}  
\usepackage{amssymb}
\usepackage{amsmath}
\usepackage{color}  

\usepackage{dsfont}

\begin{document}

\title{ Density-constrained Time-dependent Hartree-Fock-Bogoliubov method }

\author{Guillaume Scamps}
\email{gscamps@ulb.ac.jp}
\affiliation{Center for Computational Sciences, 
University of Tsukuba, Tsukuba 305-8571, Japan}
\affiliation{Institut d’Astronomie et d’Astrophysique, Universit\'e Libre de Bruxelles, Campus de la Plaine CP 226, BE-1050 Brussels, Belgium}

\author{Yukio Hashimoto}                     
\email{hashimoto.yukio.gb@u.tsukuba.ac.jp}
\affiliation{Center for Computational Sciences, 
University of Tsukuba, Tsukuba 305-8571, Japan}

\begin{abstract}
\begin{description}
\item[Background] 
 The Density-constrained Time-dependent Hartree-Fock method is currently a tool of choice to predict fusion cross-sections. However, it does not include   pairing correlations, which have been found recently to play an important role.
 
\item[Purpose]  
To describe the fusion cross-section with a method that includes the superfluidity and to understand the impact of pairing on both the fusion barrier and cross-section.

\item[Method] 
 The density-constrained method is tested first on the following reactions without pairing,  $^{16}$O+$^{16}$O and $^{40}$Ca+$^{40}$Ca. A new method is developed, the Density-constrained Time-dependent Hartree-Fock-Bogoliubov method. Using the Gogny-TDHFB code, it is applied to the reactions $^{20}$O+$^{20}$O and $^{44}$Ca+$^{44}$Ca.
 
\item[Results] The Gogny approach reproduces the experimental data well for reaction $^{16}$O+$^{16}$O and $^{40}$Ca+$^{40}$Ca. The DC-TDHFB method is coherent with the TDHFB fusion threshold. The effect of the phase-lock mechanism is shown for those reactions.

\item[Conclusions] 
 The DC-TDHFB method is a useful new tool to determine the fusion potential between superfluid systems and to deduce their fusion cross-sections.
\end{description}
\end{abstract}

\maketitle

\section{Introduction}
\label{Sec:intro}

The effect of superfluidity on the fusion reaction is not completely understood. Recently, a strong effect of the pairing gauge angle has been found using the Time-dependent Hartree-Fock-Bogoliubov theory \cite{Has16,Mag16,Sek17,Bul17,Sek17b}. It was found from those dynamical models that when both the target and the projectile are superfluid, their fusion is easier when the gauge angles of the initial nuclei are aligned and more difficult when the gauge angles are opposite.
It is expected that it will increase the fluctuations of the fusion barrier. 
To reveal empirically this effect, a systematic analysis of the  barrier distribution obtained experimentally has been done in Ref. \cite{Sca18}.
An enhancement of the barrier height of about 1 MeV for superfluid systems was reported. 
To understand and to better describe theoretically the effect of superfluidity it would be useful to have access to the Nucleus-Nucleus potential accompanied by pairing.

The Density-constrained Time-dependent Hartree-Fock (DC-TDHF) method \cite{Cus85,Uma85,Uma06,Uma06b,Uma07,Uma08,Uma09,Uma09b,Uma10,Uma12,Uma12b,Uma15,Sim13,Uma14,Ste14,Jia14,God17}  determines the Nucleus-Nucleus potential from a single Time-dependent Hartree-Fock trajectory. This potential takes into account implicitly all dynamical effects during the crossing of the barrier and then can be directly used by a coupled-channel code to predict the fusion cross-section. This approach was succesful in reproducing the fusion cross section without any adjusted parameter. Nevertheless, this approach is limited to reactions where pairing does not play a role.

In addition to the direct effect of the gauge angle, the pairing correlation can impact the shape and the deformability 
of the fragments which can then affect the fusion potential. Then to study and to predict the experimental fusion cross-section of superfluid systems it is required to develop an approach beyond the DC-TDHF method.
Continuing the long term goal of including   pairing in all the dynamical mean-field approaches \cite{Ave08,Eba10,Sca13,Ste11,Sca15,Sca17,Tan17}, the objective of the present article is to generalize the DC-TDHF method to include the Bogoliubov formalism of pairing treatment. 

The paper is organized as follows. In Sec \ref{Sec:without}, we present the DC-TDHF method  
and   apply it for the first time with a Gogny interaction to the fusion reactions $^{16}$O+$^{16}$O and $^{40}$Ca+$^{40}$Ca to reproduce the fusion cross-section.
In Sec. \ref{Sec:with}, we develop the DC-TDHFB (DC-TDHF+Bogoliubov) theory and propose a prescription for practical realistic applications. Then, this method is applied to the reactions $^{20}$O+$^{20}$O and $^{44}$Ca+$^{44}$Ca. Our conclusions are presented in Sec. \ref{sec:conclusion}.

\section{Method without pairing}
\label{Sec:without} 

The DC-TDHF method consists of two procedures: First, a TDHF evolution is calculated at an energy above the barrier.  
Second, a minimization of the energy is performed with the constraint over the density distribution in the real space.

In the second step, it is necessary to minimize the energy with a sum of Lagrange constraints,
\begin{align}
E = E_{HF} - \sum_{\bf r} \lambda({\bf r}) \left(  \rho_{{\bf r},{\bf r} } - \rho_0({\bf r} ,t)  \right). 
\end{align} 
This leads to solving the usual HF equation with an adjusted potential,
\begin{align}
\left( \hat h - \lambda({\bf r}) \right) | \varphi_i \rangle = \epsilon_i | \varphi_i \rangle.
\end{align}
This equation is solved iteratively, with a full diagonalization of the Hamiltonian. The Lagrange parameters are readjusted as
\begin{align}
\lambda^{(n+1)}({\bf r}) =  \lambda^{(n)}({\bf r}) +  (\rho_{{\bf r},{\bf r}} - \rho_0({\bf r})) \frac{a}{ \rho_{{\bf r},{\bf r}}  +d_0 }. \label{eq:adj_lamda}
\end{align}
The value of $a$~=~-0.5~MeV and $d_{0}$~=~0.5 fm$^{-3}$ have been used.
The iterative process is accelerated using the modified Broyden's method \cite{Bar08,Joh88}.

	\begin{figure}[htb]
		\begin{center}
			\includegraphics[width=  \linewidth]{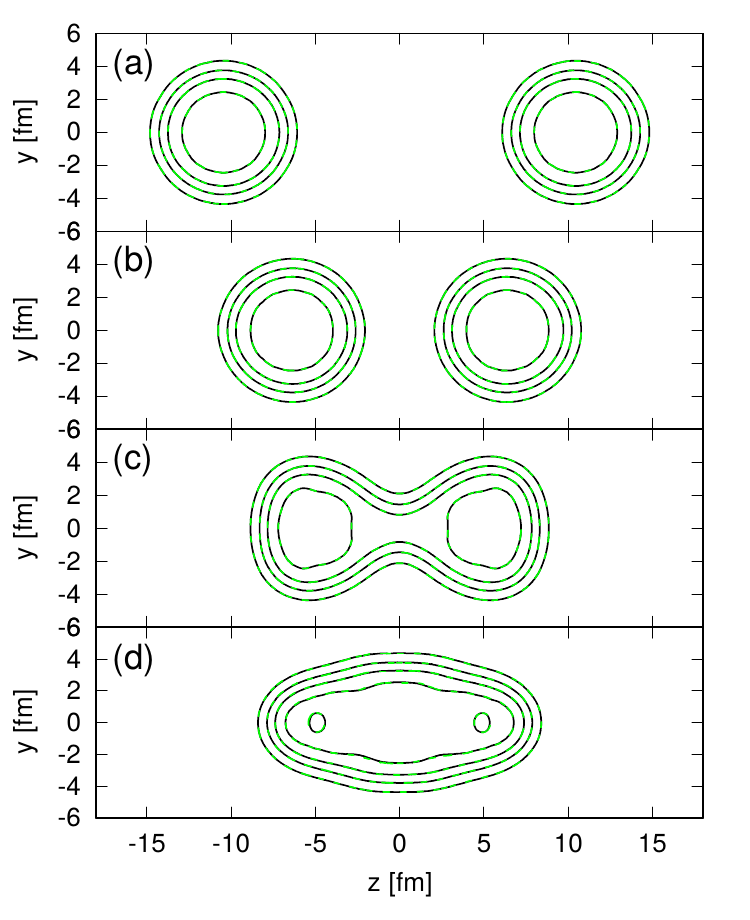}
		\end{center}
		\caption{Comparison between the TDHF (black solid line)  and DC-TDHF densities (green dashed line) as a function of time for the system $^{40}$Ca+$^{40}$Ca at center of mass energy $E_{\rm c.m.}$~=~55~MeV. The contour  lines are computed  at the densities 0.04, 0.08, 0.12 and 0.16~fm$^{-1}$. The panels (a) to (d)  correspond to the times 0, 180, 360 and  540~fm$\,$c$^{-1}$, respectively. } 
		\label{fig:fig_Ca40_E55}
	\end{figure}

We applied this method to the reactions $^{16}$O+$^{16}$O and $^{40}$Ca+$^{40}$Ca. 
The calculations have been done by making use of the Gogny-TDHFB code \cite{Has12,Has13,Has16} 
in a simplified form of a TDHF version. 
In our calculations, we used the Gogny D1S effective interaction and a hybrid basis 
of two-dimensional harmonic oscillator eigenfunctions and one-dimensional spatial grid points or mesh. 
The reaction axis (z-axis) is described in a lattice space of mesh parameter $\Delta z$~=~0.91 fm 
with $N_z$~=~23 points for the HF initialization of the fragments and $N_z$~=~46 for the dynamics  
of the two nuclei in head-on collision reactions.
The x-y plane space is described by harmonic oscillator eigenfunctions restricted to $n_x + n_y \leq {N_{\rm shell}}$.
This parameter $N_{\rm shell}$ varies in our calculation depending on the size of the fragments 
and the convergence of the results with respect to this parameter is carefully checked in the present contribution.
The harmonic oscillator parameter $\hbar \omega$ has been adjusted to minimize the energy of the initial fragments for each system and choice of  $N_{\rm shell}$. 
The time-dependent equations are solved with the Runge-Kutta method in the fourth order with a time-step 
$\Delta t$~=~0.4~fm$\,$c$^{-1}$. 
A few changes have been done in order to accelerate the TDHFB code: i)  global optimization of the code, ii) inclusion of a cut-off in the range of the Gaussian interaction, and iii) use of the finite difference method at the eighth order instead of the Lagrange basis. 
As a consequence, the numerical cost of a TDHFB calculation has been reduced to 24 hours using 32 CPUs for a $^{44}$Ca+$^{44}$Ca reaction with $N_{\rm shell}$~=~6.

While the calculations are done on the hybrid basis, the constraints are applied on the diagonal part of the density in the position basis.
In that aim, at each iteration, the local density is computed as
\begin{align}
\rho( x_i, y_j, z_k) = \sum_{n_x', n_y', n_x, n_y} & \phi^*_{n_x'}(x_i)  \phi^*_{n_y'}(x_j)  \nonumber \\
& \rho_{n_x' n_y' z_k n_x n_y z_k } \phi_{n_x}(x_i)  \phi_{n_y}(x_j), 
\end{align}
with $x_i$ and $x_j$ corresponding to the Gauss-Hermite integration points and 
$\phi^*_{n}(x)$ the harmonic-oscillator wavefunctions. 
$\lambda(\bf r)$ is then computed in that basis before the inverse transformation is applied to transfer it back on the hybrid basis.

	\begin{figure}[htb]
		\begin{center}
			\includegraphics[width=  \linewidth]{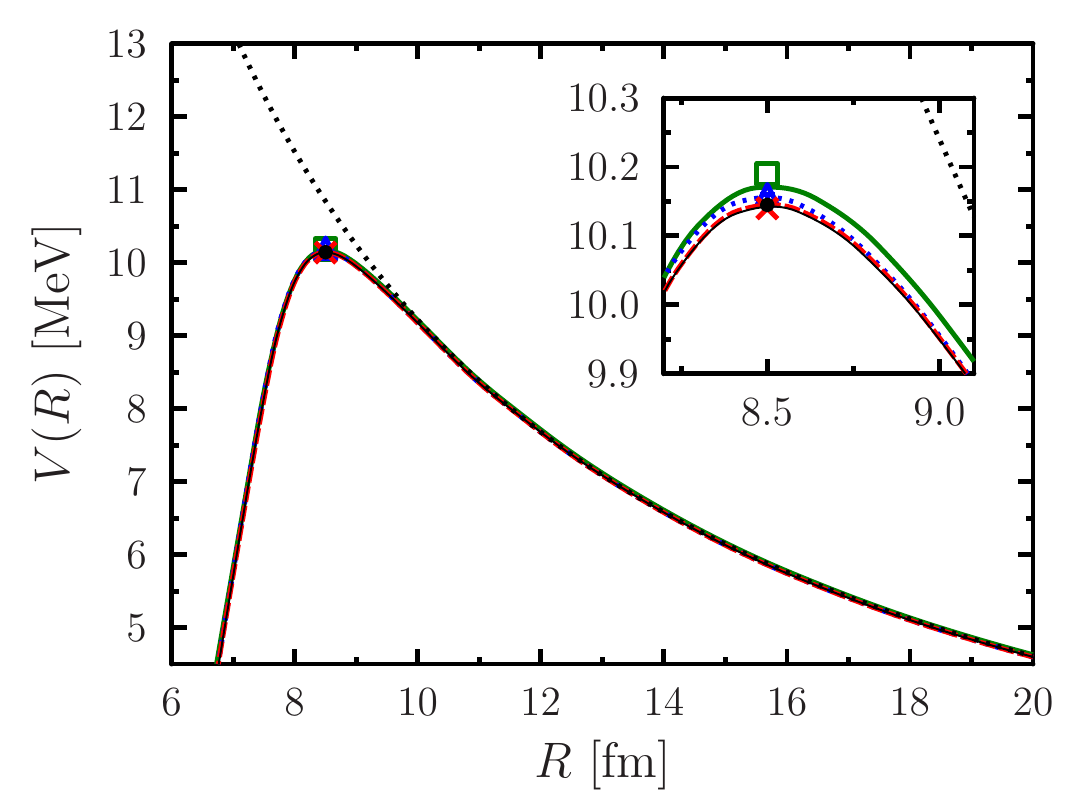}
		\end{center}
		\caption{ Nucleus-Nucleus potential obtained with the DC-TDHF method for the reaction $^{16}$O+$^{16}$O at $E_{\rm c.m.}$~=~11 MeV. The calculation is done with $N_{\rm shell}$~=~2  (green solid line), $N_{\rm shell}$~=~3 (blue dotted line),  $N_{\rm shell}$~=~4 (red dashed line), $N_{\rm shell}$~=~5 (black thin line). The position of the TDHF  threshold barrier is shown respectively by a square, triangle, cross and dot. The point-Coulomb potential is shown by a dotted line.} 
		\label{fig:pot_conv_Nshell}
	\end{figure}
	
In order to ensure the convergence of the Hartree-Fock solver with density-constrained,
the occupation number of the states has been kept constant even in the case of level-crossing at the fermi-energy. 
To that aim, we used the maximum overlap criteria \cite{Iwa94}. 
After each diagonalisation of the Hartree-Fock Hamiltonian, 
the values of the overlap between the new wave functions $|\varphi_j^{i+1}\rangle$ and the wave functions $|\varphi_j^{i}\rangle$ at the previous iteration $i$ are calculated.
The occupied wave functions are then chosen in order to maximize the sum, $ \sum_j \langle  \varphi_j^{i} | \varphi_j^{i+1} \rangle$. 

Finally,the Nucleus-Nucleus potential is obtained as
\begin{align}
	V(R)=E_{DC}(R)-E_1-E_2, \label{eq:pot}
\end{align}
where $R$ is the distance between the fragments, $E_{DC}(R)$ is the density constrained energy, and $E_{1,2}$ are the ground state energies of each fragment.

\subsection{ Application for the $^{16}$O+$^{16}$O and $^{40}$Ca+$^{40}$Ca  reactions}

A test of the method is shown on Fig.~\ref{fig:fig_Ca40_E55}. 
We see a good agreement between the initial density obtained after the TDHF evolution and the one obtained after constraints. 

\begin{figure}[htb]
		\begin{center}
			\includegraphics[width=  \linewidth]{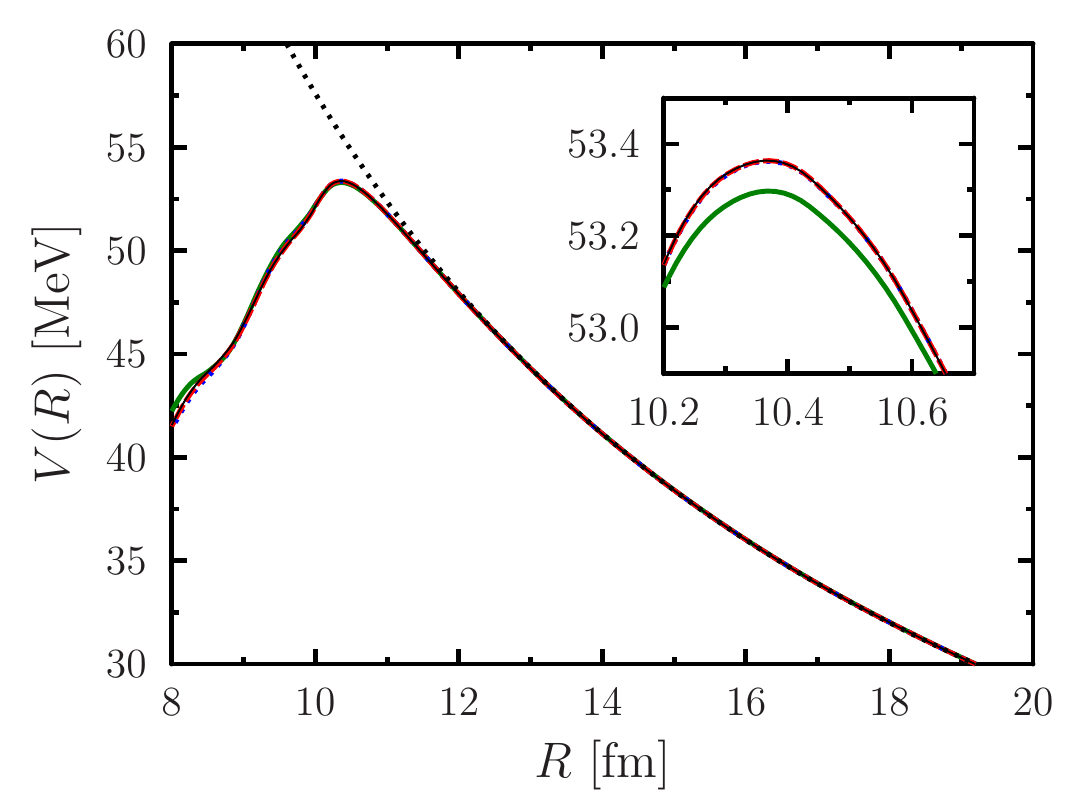}
		\end{center}
		\caption{ Nucleus-Nucleus potential obtained with the DC-TDHF method for the reaction $^{40}$Ca+$^{40}$Ca at $E_{\rm c.m.}$~=~55 MeV. The calculation is done with number of shell, $N_{\rm shell}$~=~3 (green solid line), $N_{\rm shell}$~=~4 (blue dotted line),  $N_{\rm shell}$~=~5 (red dashed line), $N_{\rm shell}$~=~6 (black thin line). The point-Coulomb potential is shown by a thin dotted line.} 
		\label{fig:pot_conv_Nshell_Ca40}
	\end{figure}

The minimal energy obtained after the convergence of the DC-TDHF calculation subtracted by  the ground state energy at an infinite distance gives the Nucleus-Nucleus potential (See Figs. \ref{fig:pot_conv_Nshell} and \ref{fig:pot_conv_Nshell_Ca40}). 
In these figures, we test the convergence of the potential with respect to the number of shells used in the hybrid basis. 
In all cases, at large distances the potential correspond well to the point-Coulomb potential showing that our DC-TDHF calculation is able to correctly cancel the collective currents. 
For the system $^{16}$O+$^{16}$O, the potential is already  well described with $N_{\rm shell}$~=~2 and no difference is seen between the results with $N_{\rm shell}$~=~4 and 5. It is also remarkable in this system that the TDHF threshold energy, the center of mass energy for which the system passes the barrier, corresponds well to the top of the DC-TDHF potential. The position of the barrier $R_b$ is determined for TDHF as the distance $R$ for which the relative impulsion is minimal. This position also matches also well with the top of the density constrained potential.

	\begin{figure}[htb]
\begin{center}
\includegraphics[width=  \linewidth]{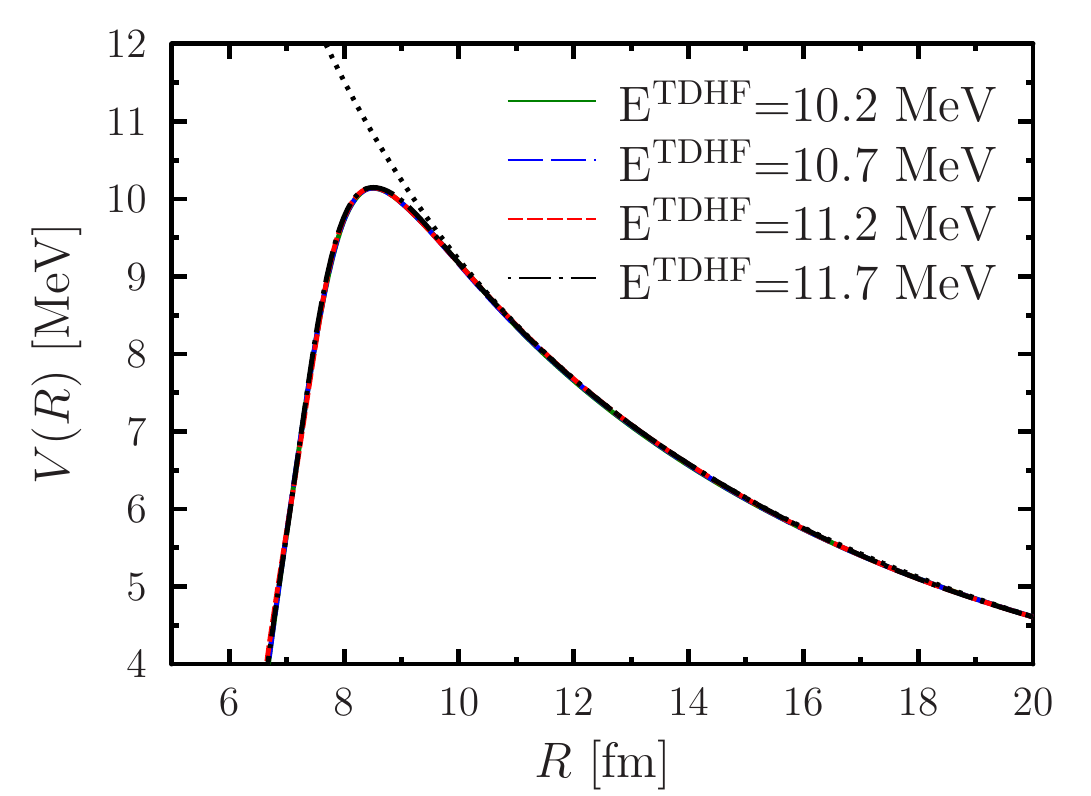}
\includegraphics[width=  \linewidth]{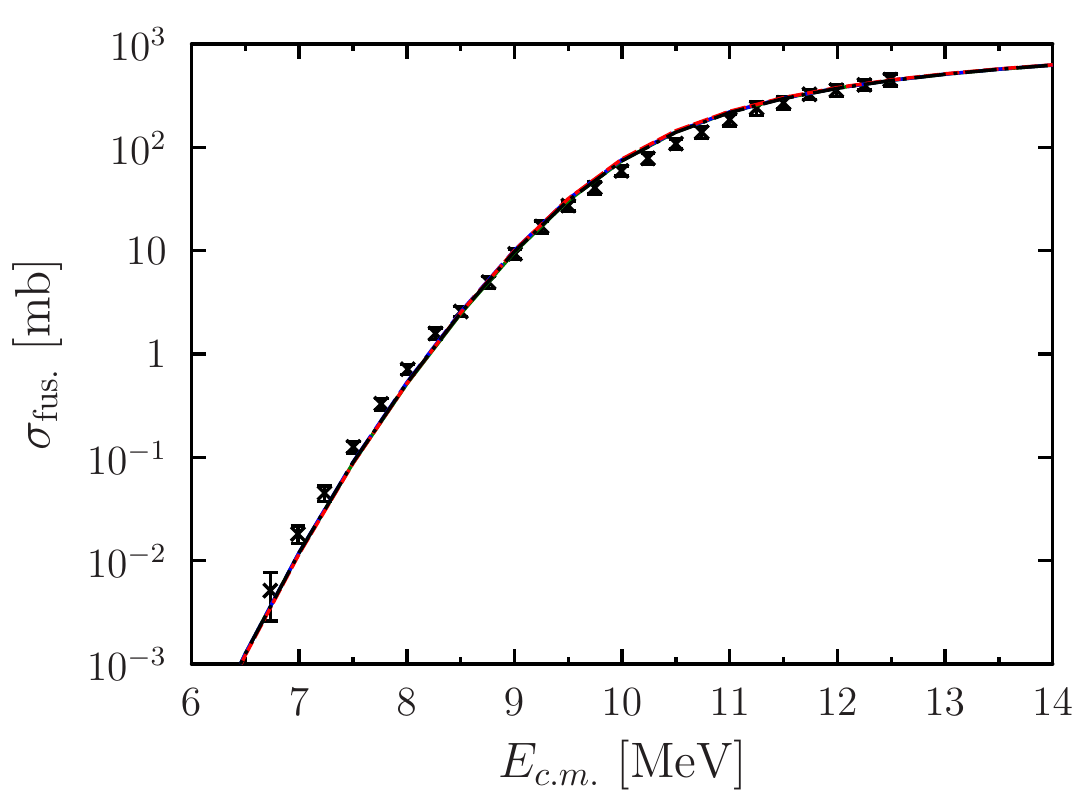}
\end{center}
\caption{Top:  Nucleus-nucleus potential obtained with the DC-TDHF theory for the reaction $^{16}$O+$^{16}$O with $N_{\rm shell}$=5 for different TDHF energies. The dotted line represents the point coulomb potential. Bottom: Fusion cross-section predicted by the  DC-TDHF method compared to the experimental values from \cite{Tho85}} 
\label{fig:pot_O16_O16}
\end{figure}

The potential for this reaction is also shown as a function of the center of mass energy in Fig.~\ref{fig:pot_O16_O16}. The potential is unchanged with a 1.5 MeV variation of the energy above the barrier.

The same numerical test has been done for the $^{40}$Ca+$^{40}$Ca system in Fig.~\ref{fig:pot_conv_Nshell_Ca40}.  
For that reaction, the results are well converged with $N_{\rm shell}$~=~4. 
Nevertheless, an important difference from the $^{16}$O+$^{16}$O case is the dependence of the potential on the energy (see Fig.~\ref{fig:pot_Ca40_Ca40}). This dependence is due to the excitation to low energy collective modes \cite{Uma14} in particular the low energy octupole mode \cite{Sim13b}.
This effect adds complexity in the interpretation of the results as we will discuss in the next section.

\subsection{Cross-section calculations}

The program {\tt CCFULL} \cite{Hag99} is used to compute the fusion cross-section from the Nucleus-Nucleus potential obtained by eq.~\eqref{eq:pot}. This potential is modified to take into account the change of the collective mass that is computed as \cite{Uma06b}
\begin{align}
   M(R) = \frac{2(E_{c.m.}-V(R))}{ \dot{R}^2 }, \label{eq:mass_Umar}
\end{align}  
where $\dot{R}$ is the relative velocity determined by the TDHF calculation. The coordinate transformation of the potential is obtained from \cite{Goe83}
\begin{align}
   d\bar{R}=\sqrt{\frac{M(R)}{\mu}} dR,
\end{align}  
where $\mu$ is the reduce mass.

\begin{figure}[htb]
\begin{center}
\includegraphics[width=  \linewidth]{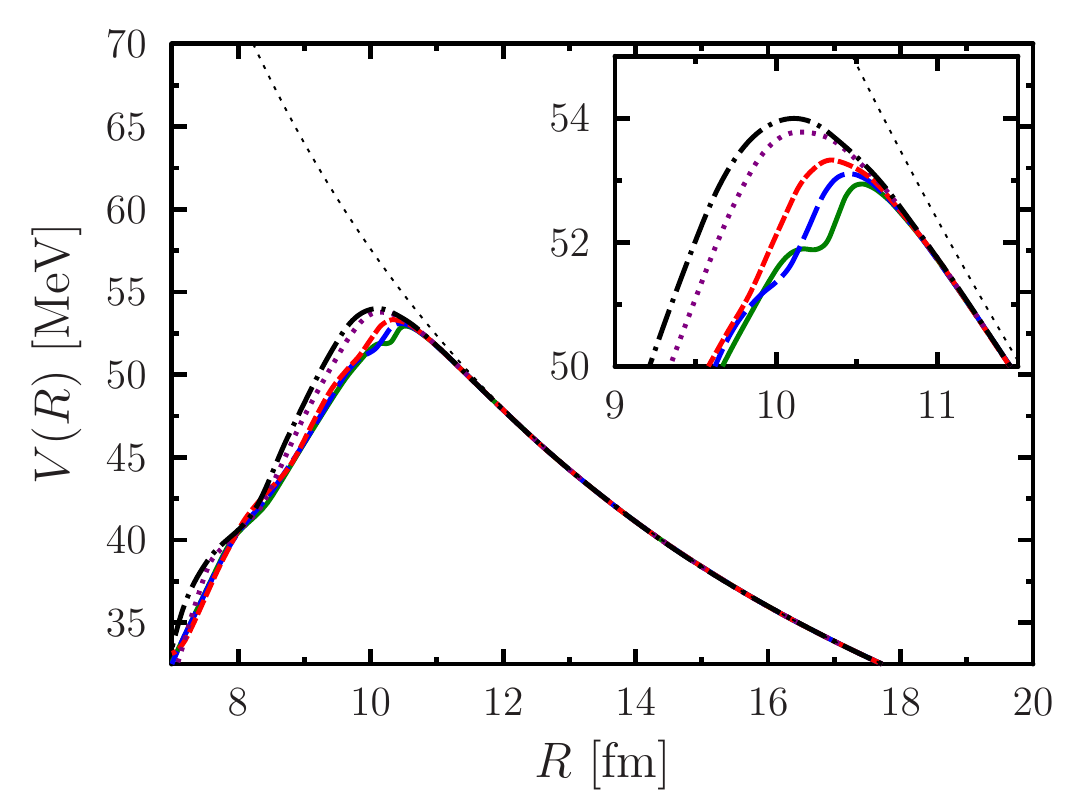}
\includegraphics[width=  \linewidth]{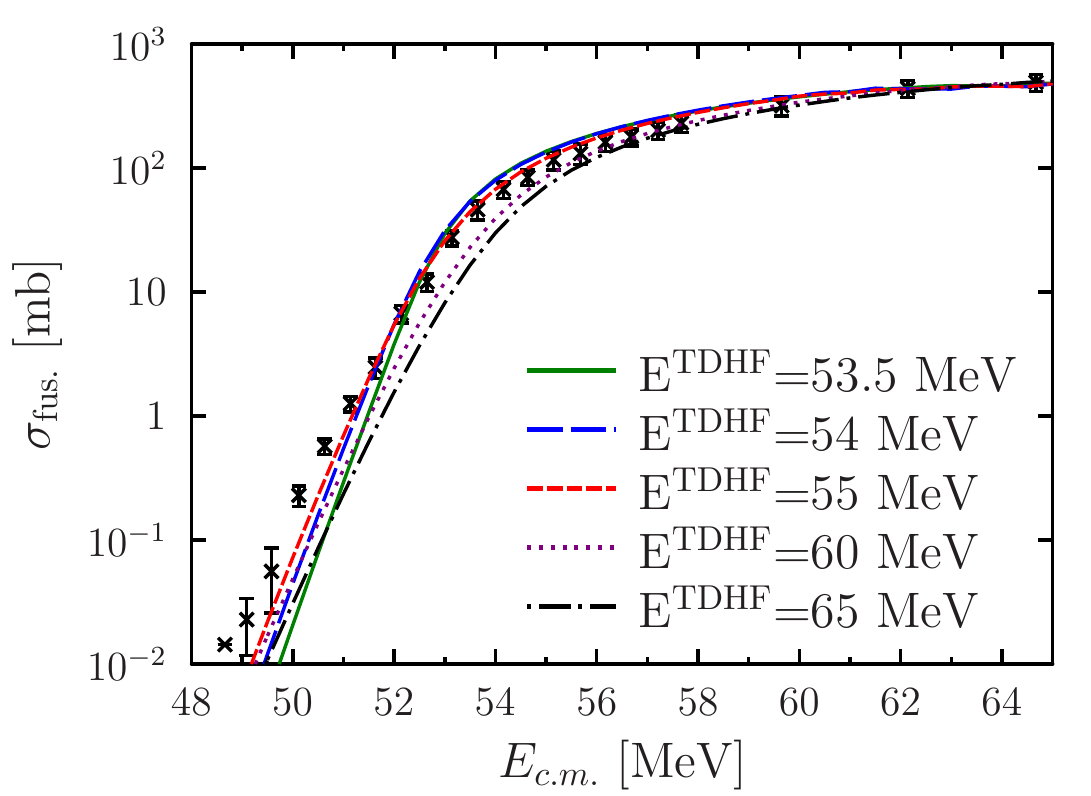}
\end{center}
\caption{Same as \ref{fig:pot_O16_O16} for the reaction $^{40}$Ca+$^{40}$Ca with $N_{\rm shell}$~=~5. The experimental data are obtained from ref. \cite{Mon12}} 
\label{fig:pot_Ca40_Ca40}
\end{figure}

The result of this calculation is shown on Fig.~\ref{fig:pot_O16_O16} for the $^{16}$O+$^{16}$O reaction. The calculation reproduces very well the experimental data and the results are independent of the center of mass energy in the TDHF simulation. This reaction was already well described via the DC-TDHF method using the Skyrme SLY4 funtional \cite{Uma12b}.

The situation is a bit different in Fig.~\ref{fig:pot_Ca40_Ca40} for the case of $^{40}$Ca+$^{40}$Ca. The fusion cross-section depends on the TDHF center-of-mass energy. To solve that issue, it was proposed to take into account an energy-dependent potential \cite{Uma14}. This method works well at energies well above the barrier but underestimates the cross-section at energies below the barrier (around $E_{\rm c.m.}$~=~50 MeV). We expect that this small discrepancy is due to the calculation of the mass by eq.~\eqref{eq:mass_Umar} which neglect the internal excitation. It would be interesting to compare, in future, applications with the prescription of \cite{Jia14}.

This first part of the article showed that the Gogny-TDHFB code using the D1S interaction is able to solve the DC-TDHF equation and to reproduce well the experimental fusion cross-section. We can now include the effect of pairing in the calculation.

\section{How to include pairing}
\label{Sec:with}
 
A first attempt to treat the pairing in the DC-TDHF method was proposed in Ref. \cite{Ste14}. The method used the BCS approximation only to determine the initial density of one of the fragment. Then the dynamical calculation was a pure TDHF calculation. Although this treatment already improved the calculation, in the present contribution we are interrested in gauge angle effects which can only be treated with the TDHFB theory, beyond the simple treatment of Ref. \cite{Ste14}. It is then necessary to develop a coherent new method, the DC-TDHFB theory.
 
The inclusion of pairing is an approximate way to treat the two-body correlation. 
Then, following the idea of the DC-TDHF method, we  apply an additional constraint on the diagonal part of the two-body density matrix,
%
\begin{align}
E = E_{HF} &- \sum_{\bf r} \lambda({\bf r}) \left(  \rho_{{\bf r},{\bf r}} - \rho_0({\bf r} ,t)  \right) \nonumber \\  &-  \sum_{\bf r, r'} \lambda^{(2)}({\bf r}, {\bf r'} ) \left(  \rho^{(2)}_{\bf rr'rr'} - \rho^{(2)}_0({\bf r,r',t})  \right), 
\end{align} 
%
where $\rho^{(2)}_0({\bf r,r',t})$ is the TDHFB diagonal part of the density at time $t$. 
The  two-body density matrix within the HFB formalism is written as
\begin{align}
 \rho^{(2)}_{\bf r_1 r_2 r_3 r_4} = \rho_{\bf r_1 r_3 }  \rho_{\bf r_2 r_4 }  - \rho_{\bf r_1 r_4 }  \rho_{\bf r_2 r_4 } + \kappa_{\bf   r_1 r_2 } \kappa^*_{\bf   r_3 r_4 }.
\end{align} 
The diagonal part can then be written as
\begin{align}
 \rho^{(2)}_{\bf r_1 r_2 r_1 r_2} = \rho_{\bf r_1 r_1 }  \rho_{\bf r_2 r_2 }  - \rho_{\bf r_1 r_2 }  \rho^*_{\bf r_1 r_2 } + \kappa_{\bf  r_1 r_2 } \kappa^*_{\bf   r_1 r_2 }.
\end{align} 

Then constraining the diagonal part of the two-body density matrix is 
equivalent to constraining the norm of all the matrix elements $\rho_{\bf r_1 r_2 }$ and $\kappa_{\bf  r_1 r_2 }$.

Because it is not possible in practice to constrain all those matrix elements in realistic calculations, we use the following prescription. In the present DC-TDHFB calculation, we apply a constraint on the diagonal element of the one-body density  $\rho({\bf r},{\bf r} )$ 
and one on the norm of all the matrix elements  $\kappa_{ij}$ where $i$ and $j$ are the labels of the hybrid basis $i<=> \{n_x,n_y,z\}$. 
Therefore, the following minimization is done:
%
\begin{align}
E = E_{HFB} &- \sum_{\bf r} \lambda({\bf r}) \left(\rho_{{\bf r},{\bf r}} - \rho_0({\bf r} ,t)    \right) \nonumber \\  &-  \sum_{i,j} \lambda^{(2)}_{ij} \left(  \kappa_{ij} \kappa^*_{ij}  - \kappa^{(0)}_{ij}(t) \kappa^{(0)*}_{ij}(t) \right). 
\label{eq:prescription_DCTDHFB}
\end{align} 
%

This prescription ensures that after the minimization 
i)  the local density is the same as that of the TDHFB state ii) the pairing energy is essentially conserved, 
iii) the minimization has enough degrees-of-freedom to remove all of the currents, 
and iv) in the case of no pairing the DC-TDHF equation is found.

In practice the adjustment for $\lambda({\bf r})$ is done with eq.~\eqref{eq:adj_lamda} and $\lambda^{(2)}_{ij}$ as
\begin{align}
\lambda^{(2) (n+1)}_{ij}  = \lambda^{(2)(n)}_{ij} +  \frac{\kappa_{ij}}{|\kappa_{ij}|}  \left( |\kappa_{ij}| - |\kappa_{ij}^{(0)}| \right) b \sqrt{|\kappa_{ij}^{(0)}|}, 
\end{align} 
with $b$ adjusted to ensure a fast convergence in each case  with a  typical value on the order of $b \simeq 10a$.

The HFB equation under density constraint becomes
\begin{align}    
\left(  
\begin{array} {cc}  
 h -\lambda - \lambda({\bf r}) &  \Delta - \lambda^{(2)}_{ij} \\  
 -\Delta^* + \lambda^{(2)*}_{ij} &  -h^* + \lambda + \lambda({\bf r})  \\  
\end{array}   
\right)
\left(  
\begin{array} {c}  
 u_{\alpha}  \\  
 v_{\alpha}   \\  
\end{array}   
\right)
 = \epsilon_{\alpha} 
\left(  
\begin{array} {c}  
 u_{\alpha}  \\  
 v_{\alpha}   \\  
\end{array}   
\right),
\end{align} 
where $\lambda$ is the chemical potential adjusted to conserve the total number of neutrons and protons. 
Note that this constraint is optional because the constraint of the density already insures the good total number of particles. 
Nevertheless, it is kept in order to accelerate the convergence of the calculation. In a similar way as the method without pairing, the iterative process is also accelerated by using the modified Broyden's method.

\subsection{ $^{20}$O+$^{20}$O reaction }

A first test is done for a reaction between light ions. An additional complexity arising with the treatment of superfluidity is the dependence with the relative initial gauge angle. 
Just   as in the case of two deformed nuclei 
where the relative orientation of the two nuclei changes the height of the barrier 
on the DC-TDHF calculations \cite{Uma06b}, 
we expect that the potential will be affected by the gauge angles \cite{Has16}.  

The system is initially set up with the method of Ref. \cite{Has16} at a relative gauge angle $\varphi$ i.e. a transformation is done on the left initial fragment  
\begin{eqnarray}
  U_{\alpha k} = e^{i \varphi}   U_{\alpha k}^{(0)}, \,\,  \quad 
  V_{\alpha k} = e^{- i \varphi} V_{\alpha k}^{(0)}.  \label{eq:def_angle}
\end{eqnarray}  
This definition of $\varphi$ differs from the one of Ref. \cite{Mag16} by a factor of 2. In this study, we restrain the range of variation of $\varphi$ from 0 to $\pi/2$ since for a symmetric system, the results will be unchanged by a transformation $\varphi \rightarrow -\varphi$  and $\varphi \rightarrow \pi-\varphi$.

\begin{figure}[htb]
\begin{center}
\includegraphics[width=  \linewidth]{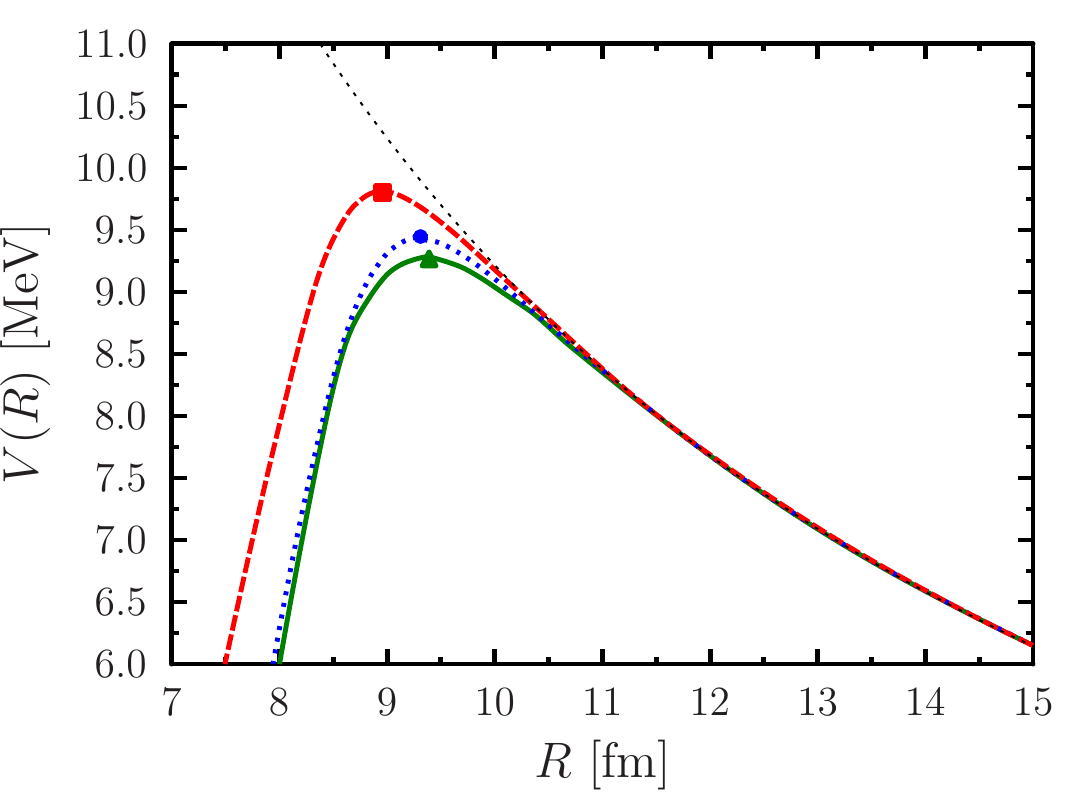}
\end{center}
\caption{$^{20}$O+$^{20}$O Nucleus-Nucleus potential  with the DC-TDHFB method  for different initial gauge angle $\varphi$ at energies just above the barrier $E_{cm}$~=~9.254 MeV, $E_{cm}$~=~9.451 MeV and $E_{cm}$~=~9.800 MeV respectively for $\varphi$~=~0 (solid green line), $\varphi$~=~$\pi/4$ (dotted blue line) and $\varphi$~=~$\pi/2$ (dashed red line). The point-Coulomb potential is shown by a thin dotted line. The threshold barrier obtained with the TDHFB method is shown by green triangle ($\varphi$~=~$\pi/4$), blue dot ($\varphi$~=~0), and red square ($\varphi$~=~$\pi/2$). Each calculation was done with $N_{\rm shell}$~=~5. } 
\label{fig:pot_O20_O20}
\end{figure}

The potential obtained by our prescription for the DC-TDHFB equation is shown on Fig.~\ref{fig:pot_O20_O20}. The three potentials show a good behavior as a function of $R$: i) at large $R$ the point-Coulomb potential is recovered ii) the position and height of the barrier are very close to the TDHFB threshold barrier. The satisfaction of these two conditions confirms the relevance of our prescription (eq.~\eqref{eq:prescription_DCTDHFB}).

Fig.~\ref{fig:pot_O20_O20}  has some similarities with Fig.~15. of Ref. \cite{Has16} obtained with the frozen density method. The highest barrier is found for $\varphi$~=~$\pi/2$ and the lowest for $\varphi$~=~$0$. Nevertheless, the $\varphi$~=~$\pi/4$ is well in the middle of the two curves with the frozen density method. In particular, the barrier height was found at 9.42, 9.6 and 9.79 MeV respectively for $\varphi$~=~0, $\pi/4$ and $\pi/2$.
In the present approach, the  $\varphi$~=~$\pi/4$ potential tends to get closer to the $\varphi$~=~$0$ curve when $R$ decreases. This decreases the difference of barrier height between $\varphi$~=~0 and $\pi/4$. This effect can be understood as a dynamical effect absent from the frozen density calculation and corresponds well to the phase locking process describe in Ref. \cite{Bul17}. Indeed, we can expect that for a relative initial phase $\varphi$~=~$\pi/4$, when the two systems are in contact the phase of the two systems will tend to align and to get closer to the $\varphi$~=~$0$ case. This effect does not appear for $\varphi$~=~$\pi/2$ for a reason of symmetry creating an unstable equilibrium. 
Indeed, the phases of the two nuclei cannot align to each other if they are exactly anti-parallel.

\begin{figure}[htb]
		\begin{center}
			\includegraphics[width=  \linewidth]{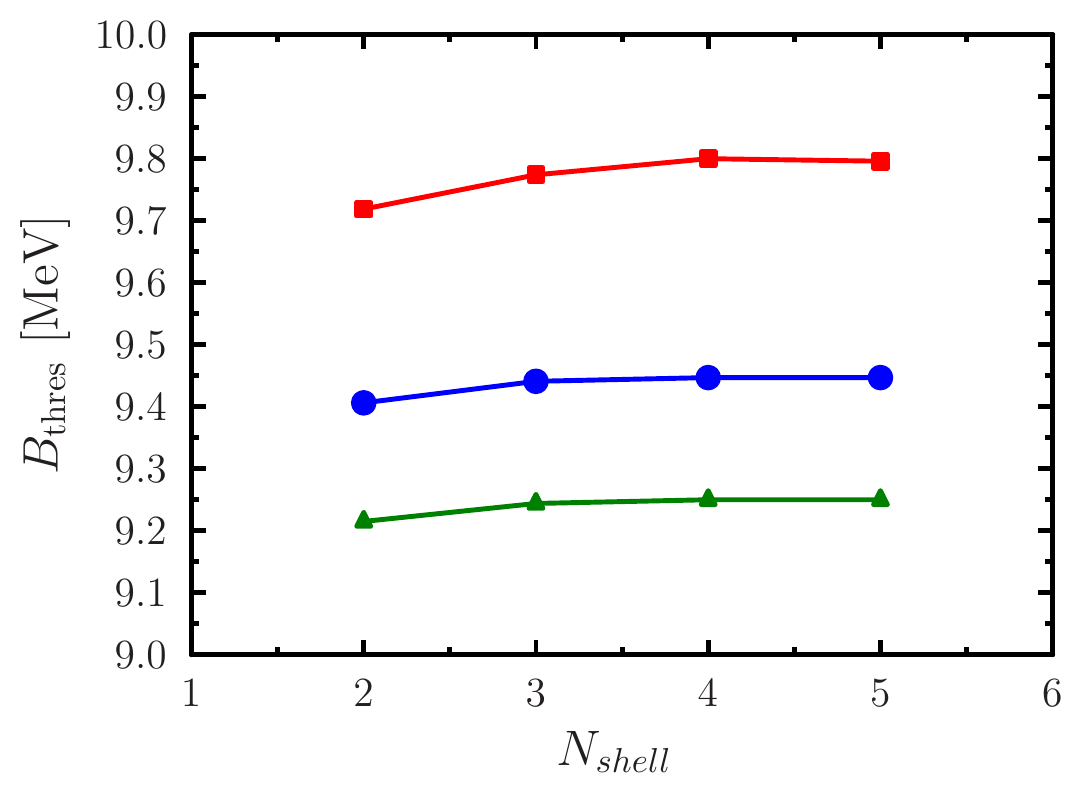}
		\end{center}
		\caption{ Threshold barrier for the $^{20}$O+$^{20}$O reaction as a function of the $N_{shell}$ value for different values of $\varphi$ (The symbols are the same as in Fig.~\ref{fig:pot_O20_O20}).} 
		\label{fig:bar_fct_N}
	\end{figure}

The convergence of the results with the size of the basis $N_{\rm shell}$ is shown on Fig.~\ref{fig:bar_fct_N}. A good approximation is obtained with  $N_{\rm shell}$~=~3 and there are no visible differences between the results with $N_{\rm shell}$~=~4 and 5. 
In particular, the splitting of the barrier is very stable with respect to the basis size.

\begin{figure}[htb]
\begin{center}
\includegraphics[width=  \linewidth]{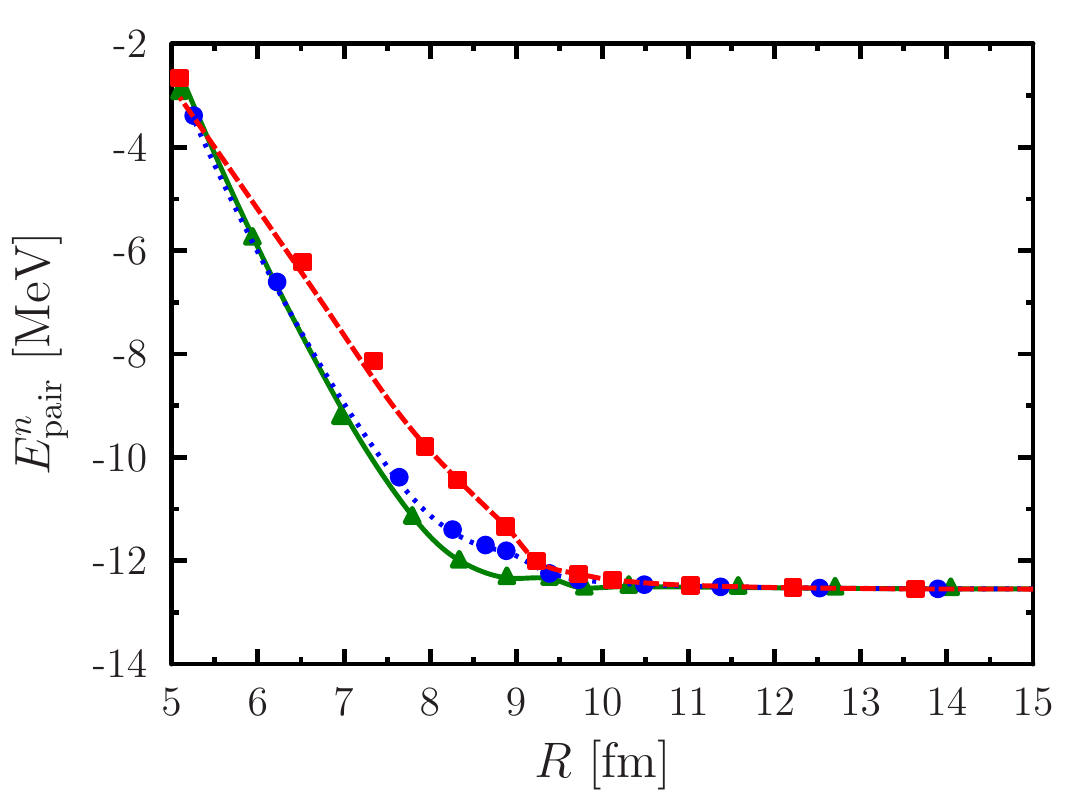}
\end{center}
\caption{ Neutron pairing energy as a fucntion of the distance between the fragments for the reaction $^{20}$O+$^{20}$O at the same center of mass energies of Fig.~\ref{fig:pot_O20_O20}. The energy obtained from the TDHFB evolution (solid green line for $\varphi$~=~0, dotted blue line for  $\varphi$~=~$\pi/4$ and dashed red line for  $\varphi$~=~$\pi/2$) is compared to the DCTDHFB pairing energy (respectively green triangles, blue dots and red squares). } 
\label{fig:Epair_O20_O20}
\end{figure}

\begin{figure}[htb]
\begin{center}
\includegraphics[width=  \linewidth]{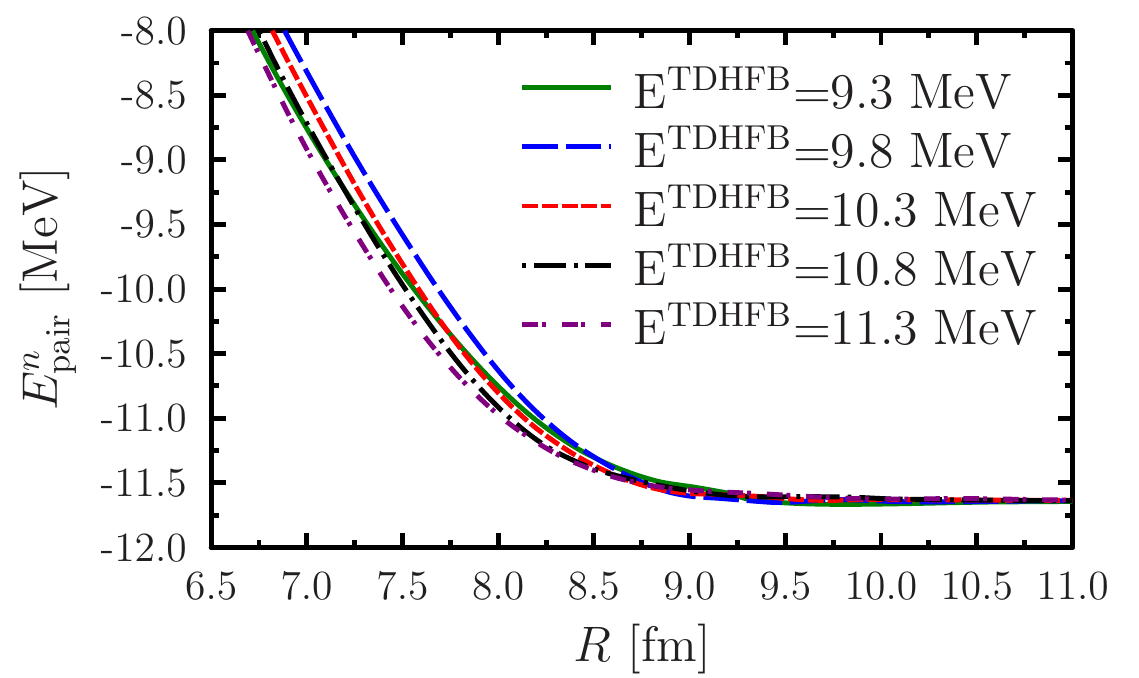}
\end{center}
\caption{ Neutron pairing energy as a fucntion of the distance between the fragments for the reaction $^{20}$O+$^{20}$O for $\varphi$~=~0 at different TDHFB center of mass energies. } 
\label{fig:Epair_O20_O20_fct_Ecm}
\end{figure}

\begin{figure}[htb]
\begin{center}
\includegraphics[width=  \linewidth]{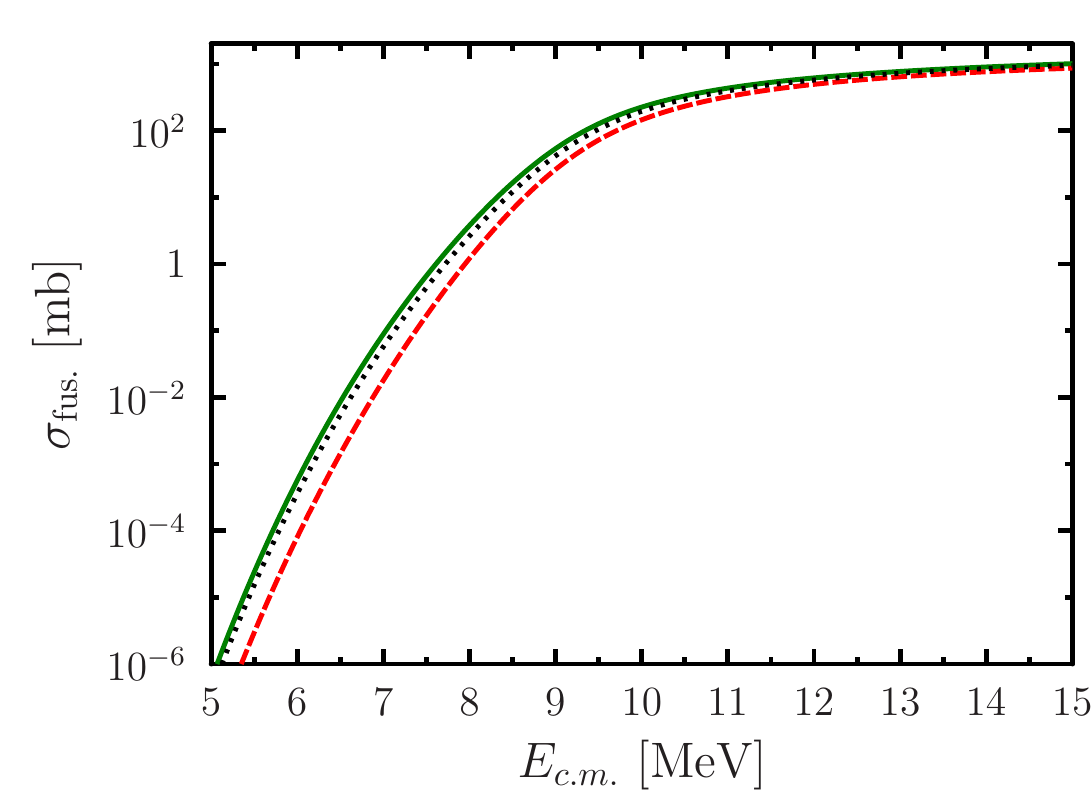}
\includegraphics[width=  \linewidth]{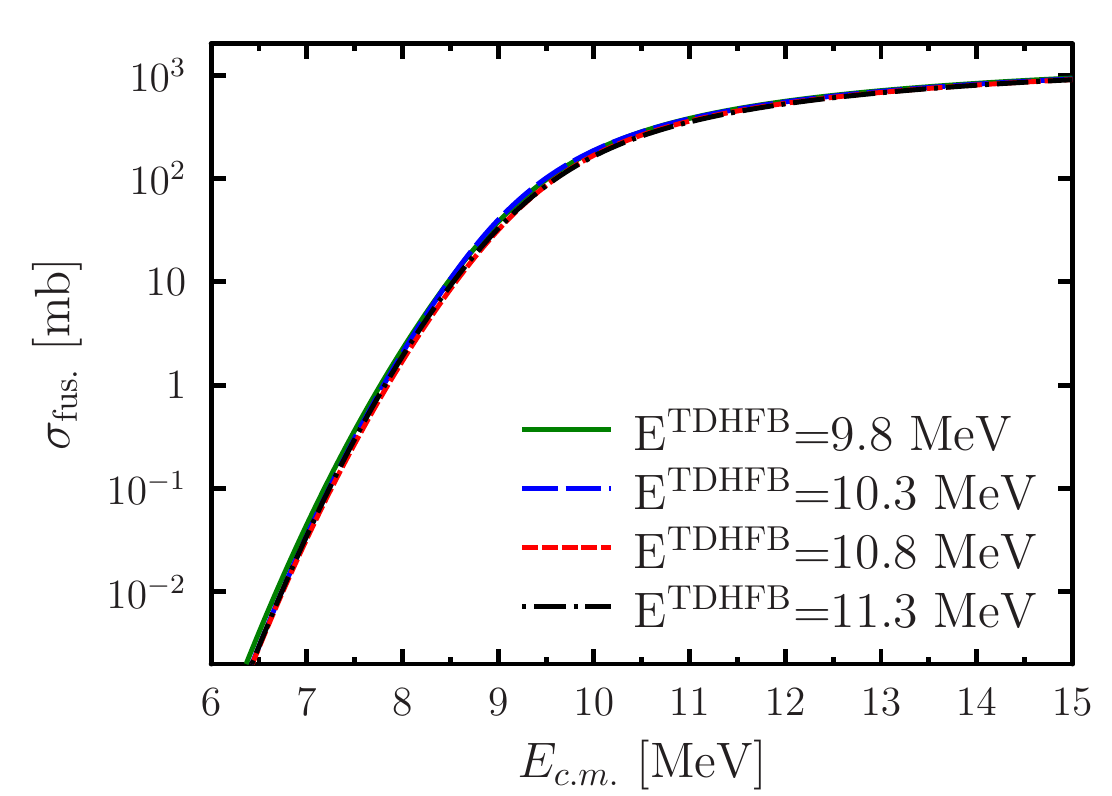}
\end{center}
\caption{Top: $^{20}$O+$^{20}$O fusion cross-section with  $\varphi$~=~0  (solid green line) and  $\varphi$~=~$\pi/2$ (dashed red line). The gauge angle averaged cross-section (eq. \eqref{eq:gauge_average}) is shown with dotted lines. The DCTDHFB potentials are evaluated at the same center of mass energies of Fig.~\ref{fig:pot_O20_O20} and with $N_{shell}$~=~5. Bottom: Dependence of the cross-section with the TDHFB energy.} 
\label{fig:cross_O20_O20}
\end{figure}

It is interesting to look at the disappearance of the superfluidity. When the two systems fuse the excitation energy tends to reduce the pairing gap. The Fig.~\ref{fig:Epair_O20_O20} shows the pairing energy disappearance during the fusion. We can see that most of the diminution of the pairing is found at distances below 8.5 fm which means after the system passes the barrier. It is remarkable that the prescription of eq. \eqref{eq:prescription_DCTDHFB} is able to correctly reproduce the evolution of the pairing energy. 

We can expect that the dissipation mechanism reducing the pairing energy will induce a strong dependence of the barrier height with the center of mass energy. However, as seen on Fig.~\ref{fig:Epair_O20_O20_fct_Ecm} the dependence of the pairing energy with the center of mass energy $E^{TDHFB}$  is weak and not trivial. Indeed, when $E^{TDHFB}$ increases, the dissipation is more important per unit of time, but also the time to cross the barrier is shorter. As a result, the fusion barrier is almost unchanged with respect to a variation of the center of mass energy. A similar behavior is found for $\varphi$~=~$\pi/4$ and  $\varphi$~=~$\pi/2$.

The fusion cross-section is obtained with the same method as for the DC-TDHF. The result is shown in Fig.~\ref{fig:cross_O20_O20} for the two extreme cases $\varphi$~=~0 and $\varphi$~=~$\pi/2$, showing the effect of the initial gauge angle. The two choices of angle modify by about one order of magnitude the fusion cross-section at energies lower than the barrier. Nevertheless, the gauge angle is not a parameter of the reaction that can be changed experimentally. Then it is necessary to restore the initial symmetry of the gauge angle using a Multi-configuration-TDHFB method \cite{Reg19,Sca17} but those options are beyond the scope of the present approach. Also, a study in a simplified model of collision between superfluid nuclei \cite{Sca17_proc,Reg18} shows that the classical restoration of the symmetry can be a good approximation. In the present case, it consists in averaging the fusion cross-section with respect to the gauge angle  
\begin{align}
	\sigma_{\rm fus.} = \frac1{2 \pi} \int_0^{2 \pi} \sigma_{\rm fus.}(\varphi) d\varphi,
	\label{eq:gauge_average}
\end{align}
using a discretization with 8 points. 
The fusion cross section by using the averaged gauge angle barrier is shown in Fig.~\ref{fig:cross_O20_O20} 
and is close to the $\varphi$~=~0 cross-section. 
This means that at low energies the component with the angle $\varphi$~=~0 dominates in the calculation, 
because the barrier with the angle $\varphi$~=~0 is lower than those with the other values of $\varphi$
and the phase locking effect makes the potential with the angle $\varphi$~=~$\pi/4$ shift toward that with angle $\varphi$~=~0. On the bottom panel of fig.~\ref{fig:cross_O20_O20}, we can see that there is only a very small dependence of the fusion cross-section on the incident TDHFB energy. The behavior being similar to the $^{16}$O+$^{16}$O reaction (Fig. \ref{fig:pot_O16_O16}), we can conclude that the superfluidity does not bring an additional energy dependence of the potential.

\subsection{  $^{44}$Ca+$^{44}$Ca reaction }

\begin{figure}[htb]
\begin{center}
\includegraphics[width=  \linewidth]{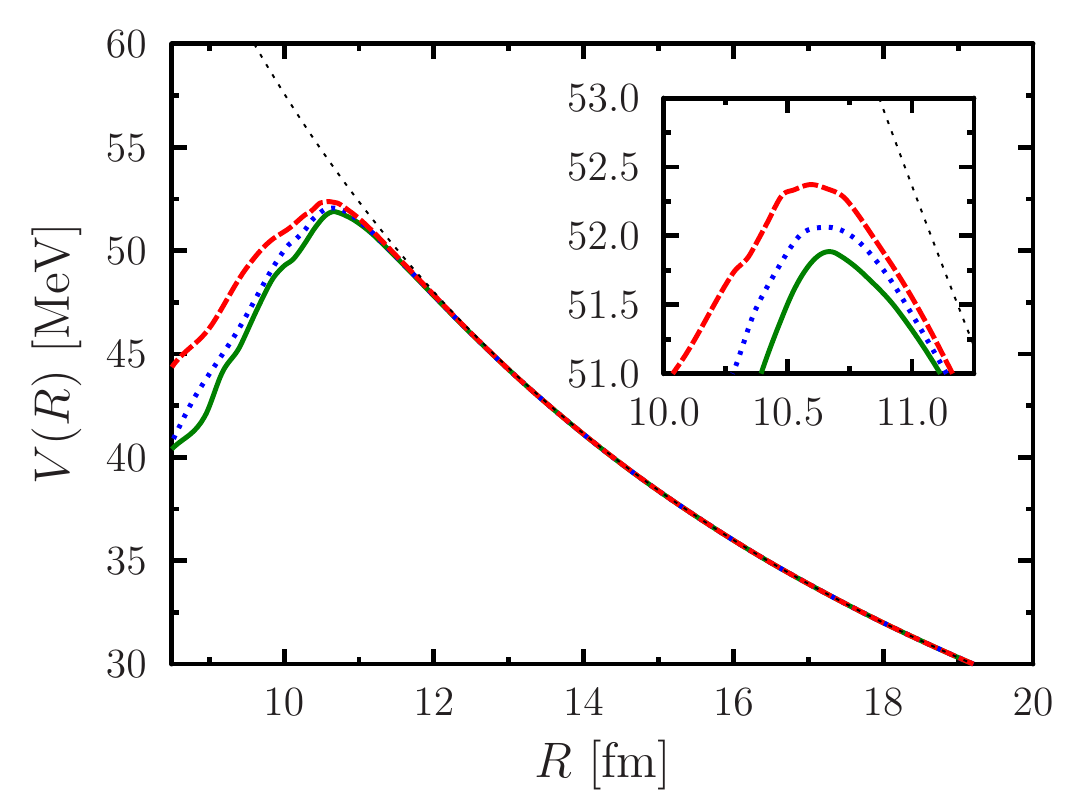}
\end{center}
\caption{$^{44}$Ca+$^{44}$Ca Nucleus-Nucleus potential obtained with the DC-TDHFB method obtained for different initial gauge angle $\varphi$ at energies just above the barrier $E_{cm}$~=~53 MeV for $\varphi$~=~0 (solid green line), $\varphi$~=~$\pi/4$ (dotted blue line) and $\varphi$~=~$\pi/2$ (dashed red line). The point-Coulomb potential is shown by the thin dotted line. Each calculation was done with $N_{\rm shell}$~=~6. } 
\label{fig:pot_Ca44}
\end{figure}

The splitting of the barrier for the  $^{44}$Ca+$^{44}$Ca reaction is shown on Fig.~\ref{fig:pot_Ca44}. The expected splitting is found,  but the phase alignment effect found in the case of $^{20}$O+$^{20}$O is not as clear here, because the potential depends on the energy like in the case of the $^{40}$Ca+$^{40}$Ca.   Because of this effect, the barrier height does not correspond to the fusion threshold which is found to be 51.63, 51.99 and 52.49 MeV respectively for $\varphi$~=~$0$, $\varphi$~=~$\pi/4$ and $\varphi$~=~$\pi/2$.

In particular, the difference between the highest and lowest barrier is 0.5 MeV in the DC-TDHFB case and 0.87 MeV for the difference of threshold energy. This suggests that dynamical effects above the barrier tend to reduce the effect of the gauge angle.

\section{Summary} 
\label{sec:conclusion}

In this article, we developed a new method to determine the Nucleus-Nucleus potential between superfluid systems. The method has been first applied to non-superfluid systems to test  the Gogny-TDHFB code and the hybrid basis. Careful tests of the convergence with respect to the basis have been done in order to show the efficiency of the hybrid basis. Despite the fact that this basis restrains the calculation to head-on collisions, it is very useful to reduce the calculation time while keeping the results unaltered.

The proposed DC-TDHFB method gives the expected splitting of the potential due to the relative gauge angle. Using that method, it has been shown that i) the phase locking process has an important effect on the Nucleus-Nucleus potential, ii) the gauge angle modifies the fusion cross-section, and iii) dynamical effects above the barrier reduce the effect of the gauge angle.

For simplicity reason, the present applications have been limited to the symmetric system  involving nuclei that are spherical in there ground-state. We plan to apply this method to other reactions for which there is experimental data such as $^{64}$Ni+$^{64}$Ni, $^{18}$O+$^{16}$O and $^{40}$Ca+$^{44}$Ca.
It will be also interesting in the future, to use the DC-TDHFB method to investigate the fission \cite{Tan15} and quasi-fission \cite{Sek19,Sek19b} processes where it can be useful to determine the potential on a dynamical trajectory.

\begin{acknowledgments}

We thank D. Lacroix, D. Regnier, T. Nakatsukasa, C. Simenel, A. Bulgac, P. Magierski and K. Sekizawa for interesting and useful discussions and E. Olsen for his careful reading of the manuscript. The calculations have been done using the COMA system at the CCS in University of Tsukuba supported by the HPCI Systems Research Projects (Project ID hp180041).
This work was supported by the Fonds de la Recherche Scientifique - FNRS and the Fonds Wetenschappelijk Onderzoek - Vlaanderen (FWO) under the EOS Project nr O022818F.\\
\end{acknowledgments}


\begin{thebibliography}{99}

\bibitem{Has16} Y. Hashimoto, and G. Scamps, Phys. Rev. C {\bf 94}, 014610 (2016).

\bibitem{Mag16} P. Magierski, K. Sekizawa, and G. Wlazlowski, Phys. Rev. Lett. {\bf 119}, 042501 (2017).

\bibitem{Sek17}  K. Sekizawa, P. Magierski,  and G. Wlazlowski, PoS(INPC2016), 214 (2017).

\bibitem{Bul17} A. Bulgac and S. Jin, Phys. Rev.  Let.,  {\bf 119},  052501 (2017).

\bibitem{Sek17b} K. Sekizawa, P. Magierski,  and G. Wlazlowski, EPJ Web Conf.  {\bf 163}, 51 (2017).




\bibitem{Sca18} G. Scamps, Phys. Rev. C {\bf 97}, 044611 (2018).


\bibitem{Cus85} R. Y. Cusson, P.-G. Reinhard, M. R. Strayer, J. A. Maruhn, and W. Greiner , Z. Phys. A {\bf 320}, 475 (1985).

\bibitem{Uma85} A. S. Umar, M. R. Strayer, R. Y. Cusson, P.-G. Reinhard, and D. A. Bromley, Phys. Rev. C {\bf 32}, 172 (1985).

\bibitem{Uma06} A. S. Umar and V. E. Oberacker,  Phys. Rev. C {\bf 74}, 021601(R) (2006).

\bibitem{Uma06b} A. S. Umar and V. E. Oberacker,  Phys. Rev. C {\bf 74}, 061601(R) (2006).

\bibitem{Uma07} A.S. Umar, V.E. Oberacker, Phys. Rev. C  {\bf 76}, 014614 (2007).

\bibitem{Uma08} A.S. Umar, V.E. Oberacker, Phys. Rev. C {\bf 77}, 064605 (2008).

\bibitem{Uma09} A.S. Umar and V.E. Oberacker, Eur. Phys. J. A {\bf 39}, 243 (2009).

\bibitem{Uma09b} A.S. Umar and V.E. Oberacker, J. A. Maruhn, and P.-G. Reinhard, AIP Conf. Proc. {\bf 1165}, 383 (2009).

\bibitem{Uma10} A. S. Umar, J. A. Maruhn, N. Itagaki, and V. E. Oberacker, Phys. Rev. Lett. {\bf 104}, 212503 (2010).

\bibitem{Uma12} A.S. Umar, V.E. Oberacker, R. Keser, J.A. Maruhn and P.-G. Reinhard, AIP Conf. Proc. {\bf 1491}, 369 (2012).

\bibitem{Uma12b} A. S. Umar, V. E. Oberacker, and C. J. Horowitz, Phys. Rev. C {\bf 85}, 055801 (2012).

\bibitem{Uma15} A. S. Umar, V. E. Oberacker, Nucl. Phys. A {\bf 944}, 238-256 (2015).

\bibitem{Sim13} C. Simenel, R. Keser, A. S. Umar, and V. E. Oberacker, Phys. Rev. C {\bf 88}, 024617 (2013).

\bibitem{Uma14} A. S. Umar, C. Simenel, and V. E. Oberacker, Phys. Rev. C {\bf 89}, 034611 (2014).

\bibitem{Ste14} T. K. Steinbach, J. Vadas, J. Schmidt, C. Haycraft, S. Hudan, R. T. deSouza, L. T. Baby, S. A. Kuvin, I. Wiedenh\"over, A. S. Umar, and V. E. Oberacker, Phys. Rev. C {\bf 90}, 041603(R), 2014.

\bibitem{Jia14}  X. Jiang, J. A. Maruhn and S. Yan, Phys. Rev. C {\bf 90}, 064618 (2014).

 \bibitem{God17} K. Godbey, A. S. Umar, and C. Simenel, Phys. Rev. C {\bf 95}, 011601(R) (2017).

\bibitem{Ave08} B. Avez, C. Simenel, and Ph. Chomaz, Phys. Rev. C {\bf 78}, 044318 (2008).

\bibitem{Eba10} {S. Ebata, T. Nakatsukasa, T. Inakura, K. Yoshida, Y. Hashimoto, and K. Yabana, Phys. Rev. C {\bf 82}, 034306  (2010).}

\bibitem{Sca13} G. Scamps, and D. Lacroix, Phys. Rev. C {\bf 87}, 014605 (2013).

\bibitem{Ste11} I. Stetcu, A. Bulgac, P. Magierski, and K. J. Roche, Phys. Rev. C {\bf 84}, 051309(R) (2011).

\bibitem{Sca15} G. Scamps, C. Simenel, and D. Lacroix, Phys. Rev. C  {\bf 92}, 011602(R) (2015).

\bibitem{Sca17} G. Scamps and Y. Hashimoto, Phys. Rev. C {\bf  96},  031602(R) (2017). 

\bibitem{Tan17} Y. Tanimura, D. Lacroix, and S. Ayik, Phys. Rev. Lett.  {\bf 118}, 152501 (2017).

\bibitem{Bar08} A. Baran, A. Bulgac, M. Forbes, G. Hagen, W. Nazarewicz, N. Schunck, and M. Stoitsov, Phys. Rev. C {\bf 78}, 014318 (2008).

\bibitem{Joh88} D. D. Johnson, Phys. Rev. B {\bf 38}, 12807 (1988).

\bibitem{Has12} Y. Hashimoto, Eur. Phys. J. A {\bf 48}, 1 (2012).

\bibitem{Has13} Y. Hashimoto, Phys. Rev. C {\bf 88}, 034307 (2013).

\bibitem{Iwa94} K. Iwasawa F. Sakata Y. Hashimoto, and J. Terasaki, Prog. Theo. phys. {\bf 92}, 1119 (1994). 

\bibitem{Sim13b} C. Simenel, M. Dasgupta, D. J. Hinde, and E. Williams, Phys. Rev. C {\bf 88}, 064604 (2013).

\bibitem{Hag99} K. Hagino, N. Rowley, and A.T. Kruppa, Comp.  Phys. Comm. {\bf123}, 143 (1999). 


\bibitem{Goe83} K. Goeke, F. Gr\"ummer, P.-G. Reinhard, Ann. Phys. {\bf 150}, 504 (1983).

\bibitem{Tho85} J. Thomas, Y. T. Chen, S. Hinds, K. Langanke, D. Meredith, M. Olson, and C. A. Barnes, Phys. Rev. C {\bf 31}, 1980(R) (1985).

\bibitem{Mon12} G. Montagnoli, A. M. Stefanini, C. L. Jiang, H. Esbensen, L. Corradi, S. Courtin, E. Fioretto, A. Goasduff, F. Haas, A. F. Kifle et al., Phys. Rev. C {\bf 85}, 024607 (2012).

\bibitem{Reg19}  D. Regnier, D. Lacroix,	arXiv:1902.06491 [nucl-th].

\bibitem{Sca17_proc} G. Scamps and Y. Hashimoto, EPJ Web Conf.  {\bf 163}, 49 (2017).

\bibitem{Reg18}  D. Regnier, D. Lacroix, G. Scamps, and Y. Hashimoto, Phys. Rev. C {\bf 97}, 034627 (2018).

\bibitem{Tan15} Y. Tanimura, D. Lacroix, G. Scamps, Phys. Rev. C 92, 034601 (2015) 

\bibitem{Sek19} K. Sekizawa, Frontiers in Physics {\bf 7}, 20 (2019).

\bibitem{Sek19b} K. Sekizawa, K. Hagino,  Phys. Rev. C {\bf 99}, 051602(R) (2019).


\end{thebibliography}
\end{document}